\documentstyle[iopconf1,epsfig,rotating]{article}

\begin{document}

\newcommand{\lsim}   {\mathrel{\mathop{\kern 0pt \rlap
  {\raise.2ex\hbox{$<$}}}
  \lower.9ex\hbox{\kern-.190em $\sim$}}}
\newcommand{\gsim}   {\mathrel{\mathop{\kern 0pt \rlap
  {\raise.2ex\hbox{$>$}}}
  \lower.9ex\hbox{\kern-.190em $\sim$}}}
\def\be{\begin{equation}}
\def\ee{\end{equation}}
\def\ee{\end{equation}}
\def\ba{\begin{eqnarray}}
\def\ea{\end{eqnarray}}
\def\ap{\approx}
\def\eps{\varepsilon}

\hoffset=2cm
\voffset=2cm

\title{%
\vbox to 0pt{{\ }\par\vskip-2.5cm \noindent
\normalsize Contribution to the Proceedings of
{\it Beyond the Desert}, Schlo{\ss} Ringberg, Germany
June 8--14, 1997.\vfil}
\vskip -24pt
Are strongly magnetized degenerate stars cooling by axion emission?}

\author{M. Kachelrie{\ss}}

\affil{INFN, Lab. Naz. del Gran Sasso, I--67010 Assergi (AQ), Italy}

\beginabstract
We considered recently as a new axion production mechanism
the process $e^- \to e^- +a$ in a strong magnetic field $B$.
Requiring that for a strongly magnetized neutron star the axion 
luminosity is smaller than the neutrino luminosity we obtained
the bound $g_{ae} \lsim  10^{-10}$ for the axion 
electron coupling constant. This limit is considerably weaker than the
bound derived earlier by Borisov and Grishina using the same 
method. Applying a similar argument to magnetic white dwarf stars we obtained
$g_{ae}\lsim 9 \cdot 10^{-13} (T/10^7 {\rm K})^{5/4} 
(B/10^{10} {\rm G})^{-2}$,
where $T$ is the internal temperature of the white dwarf. Here we
note that
the observed lack of magnetized white dwarfs with low-temperature in
the galactic disc
could also be interpreted as a signature of axion emission.
Moreover, we speculate that axion emission could explain why the
putative galactic halo population of white dwarfs is so dim.  
\endabstract

\section{Introduction} 

The axion $a$ is a pseudoscalar boson introduced by Peccei and Quinn
to solve  the strong CP-problem \cite{pe77}. 
Although the naturalness of this solution can be criticized from a more
fundamental point of view, the axion is still generally considered as
the best motivated cold dark matter particle apart from the neutralino. 
Moreover, considerable experimental efforts are
dedicated to axion searches.

Here we report on the calculation of a new axion production process,
namely the emission of axions by electrons in the magnetic field $B$ of a 
strongly magnetized neutron or white dwarf  (WD) star, in Ref. 
\cite{ka97}. We
derived a new limit for the axion electron coupling constant $g_{ae}$, but
noted also that the observations of magnetized WDs in
the galactic disc
show at low-temperature a lack compared to non-magnetized WDs. 
This could be interpreted as a possible signature of axion emission.
Furthermore, we speculate that a substantial fraction of the galactic dark
matter halo seen by the MACHO and EROS experiments consists of
magnetized WDs.
Since the cooling time of a magnetized WD is drastically
reduced by axion emission, 
this could explain the non-observation of halo WDs in optical
searches
\cite{suche}.

\section{Axion cyclotron emissivity from degenerated electrons}

The solutions of the Dirac equation for electrons in an external
homogenous, magnetic field are given by Landau states.
Using the gauge $A=(0,0,Bx,0)$ they can be characterized by the
Landau quantum number $N=0,1,\ldots$, the $y$ and the $z$ components
of the momentum $p$, and the eigenvalue $\tau=\pm 1$ of a suitable polarization
operator \cite{so68}. 
Axion cyclotron emission is the transition of an electron from an
excited Landau level $n$ into a level $n'<n$ thereby emitting an axion
$a$ with momentum $k=(\omega, \vec k)$. 

The luminosity ${\cal L}_a$
emitted by ${\cal N}$ electrons occupying the volume
$V$ due to this process is given by
\begin{equation}    \label{L}
 {\cal L}_a = 
     \lim_{T \to \infty }\, \frac{1}{T} \sum_{\lambda,\lambda^\prime} 
     \sum_{\vec k}   \omega |S_{\pm}|^2  \: {\cal S} \,,
\end{equation}
where $S$ is the $S$-matrix element of the process $e^- \to e^- +a$,
the summation index $\lambda$ indicates the set of quantum numbers
$\lambda=\{ n,\tau, p_y, p_z \}$,
\begin{equation}
 {\cal S} = f (E)\left[ 1- f(E') \right] 
\end{equation}
and $f(E)$ are Fermi-Dirac distributions functions. As it is
characteristic for processes in strong magnetic fields, the $S$-matrix
element consists essentially of Laguerre functions $I$,
\begin{equation}
 I_{n^\prime,n}(\kappa)= \sqrt{\frac{n^\prime !}{n !}} \;
  \kappa^{(n-n^\prime)/2} \, e^{-\kappa/2} \, 
  L_{n^\prime}^{n -n^\prime}(\kappa) \,.
\end{equation}
The argument of the $I$-functions is given by
$\kappa= (k\sin\theta)^2 / (2eB)$, $\theta$ is the
angle between $\vec B$ and $\vec k$, and  $L_n^{n'} (x)$ are
Laguerre polynomials.
The numerical evaluation of the terms  in Eq. (1) 
becomes  cumbersome already for moderate $n$. Therefore 
we take advantage of the degeneracy of the electron gas inside inside a
neutron or white dwarf star and employ an approximation commonly used in
calculations of neutrino emission rates: Transitions between different
Landau states are only possible, if the states are lying inside the shell
$[E_F -T,E_F+T]$, where $E_F$ is the Fermi energy of the electrons. 
In this approximation, the axion cyclotron
emissivity $\varepsilon_a = {\cal L}_a /V$ is given by
\begin{equation}    \label{eps,approx}
 \varepsilon_a = \frac{eB}{2\pi^2} \:  
                 \sum_{n=1}^{n_{\rm max}}\sum_{n' < n}\sum_{\tau = \pm}
                 \int_0^\pi d\theta \sin\theta  \: \frac{E_0}{p_z}
                 \left(\frac{E_0}{E - p_z \cos\theta} \right)^2
                 \frac{d\Gamma_{\pm 0}^{n \to n'}(\vec k)}{d\theta_0} \:
                 \frac{\omega^2}{e^{\beta\omega}-1} \,.     
\end{equation}
Here, $d\Gamma_{0}$ is the differential decay width of an electron in its
rest frame as given in Ref. \cite{ka97} and
$n_{\rm max} = {\rm int} \, [(E_{F}^2 -m_e^2)/(2eB)]$.
Note that Eq.~(\ref{eps,approx}) is valid for arbitrary magnetic field
strengths. 

Although the two infinite sums over $n$ and $n'$ have now been
replaced by
finite sums, for low magnetic field strengths or high densities it is
still necessary to compute Laguerre polynomials with high index. 
This can be avoided by the use of a Bessel function approximation   
in the semiclassical, ultrarelativistic case and by the use of
the $\kappa\to 0$ limit of the $I$-functions in the classical case.

Our numerical evaluation of Eq.~(\ref{eps,approx}) confirmed  
the semi-classical limit valid for $E\gg m_e$ and 
$B\ll B_{\rm cr}=m_e^2 /e\ap 4.41 \times 10^{13}\:$Gauss,
{\it i.e.\/} for neutron stars with not too high magnetic fields, 
derived in Ref. \cite{bo94} by Borisov and Grishina. However,
applying axion cyclotron emission as an additional cooling 
mechanism to neutron stars and requiring that the axion luminosity is
smaller than the total neutrino luminosity, 
we could constrain the axion electron coupling constant only to
$g_{ae} \lsim {\cal O} (10^{-10})$. This bound is three orders of
magnitude  weaker than the bound derived in Ref. \cite{bo94}
considering the same process. The reason for this discrepancy  is that
the authors of Ref. \cite{bo94} derived their bound by requiring 
that the emissivity due to the process $e^- \to e^- +a$ is smaller 
than due to neutrino cyclotron emission $e^- \to e^- + \nu + \bar\nu$
instead of comparing $\eps_a$ with the {\em total\/} emissivity of a
neutron star.

\begin{center}
\begin{figure}
\begin{minipage}[t]{3in}
\setlength{\unitlength}{1.0in}
\begin{picture}(6,0)
 \epsfig{file={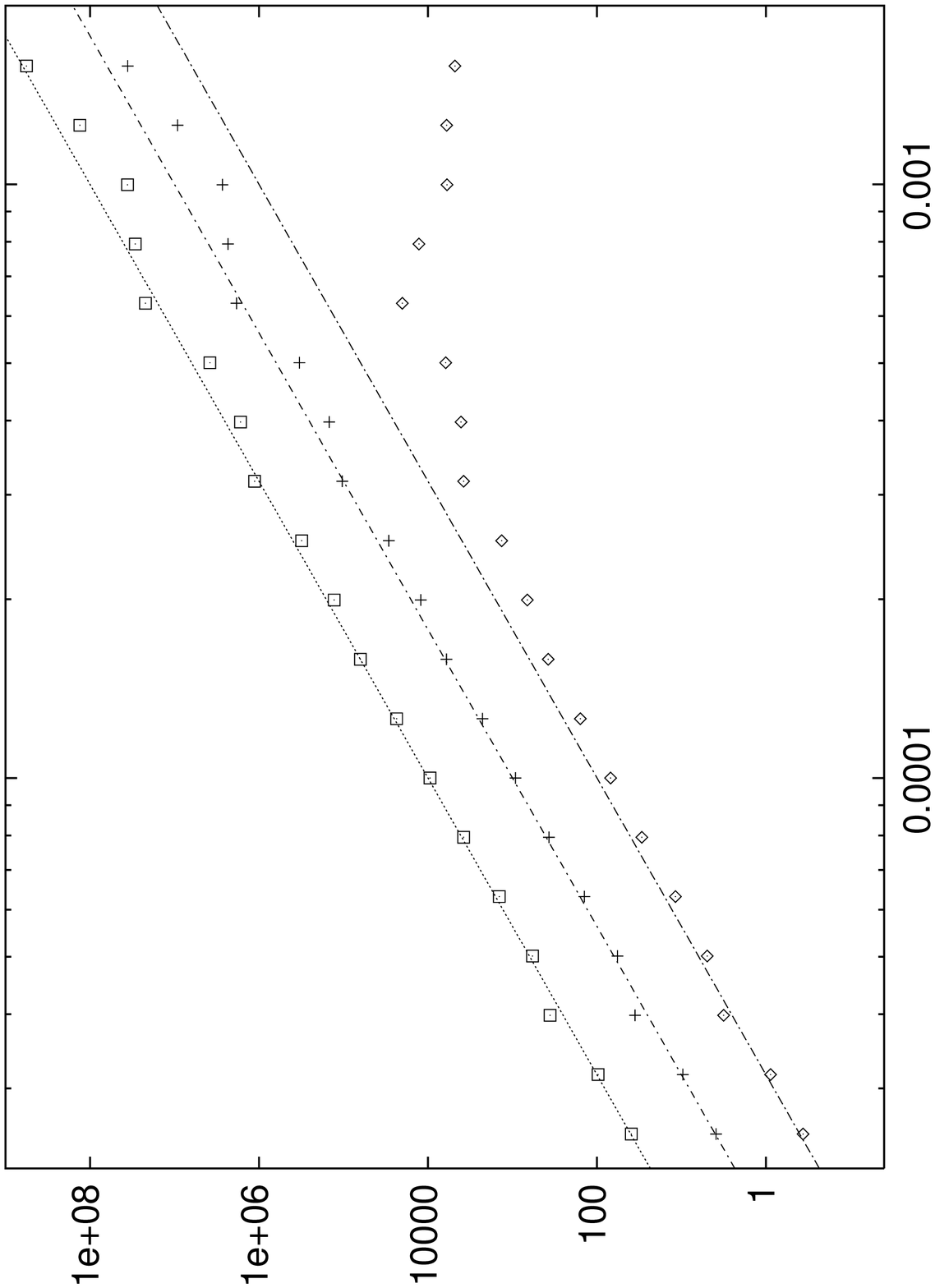}, height=3.0in,angle=270}
 \put (-2.4,-1.1){{\footnotesize $10^8$K}}
 \put (-2.4,-1.8){{\footnotesize $10^6$K}}
  \put (-1.5,-2.2){{\footnotesize $B/B_{\rm cr}$}}
 \put(-3.6,-1.1)
    {\footnotesize
       $\tilde\varepsilon_a / \frac{{\rm erg}}{{\rm cm}^3{\rm s}}$}
\end{picture}
\end{minipage}
\vskip2.3in
\caption{\label{wd}
   Axion emissivity $\tilde\varepsilon_a$ due to $e^- \to e^- + a$
  as function of $B/B_{\rm cr}$ for
  $n_e = 10^{26}$cm$^{-3}$ and $T=10^8$K, $T=10^7$K, $T=10^6$K
  (lines: fit function Eq.~(\ref{fit}), points: exact results).}
\end{figure}
\end{center}

We now apply the same line of arguments to magnetic WDs.
In Fig.~\ref{wd}, $\tilde\varepsilon_a = \varepsilon_a / \alpha_{1{\rm eV}}$
is shown for the electron density $n_e =10^{26}$cm$^{-3}$
and the temperatures $T=10^6$K, $T=10^7$K and $T=10^8$K, respectively,
together with the fit function 
\begin{equation}  \label{fit}
 \tilde\varepsilon_a^{\rm fit} = 2.6 \:\frac{\rm{erg}}{{\rm cm}^3{\rm s}} \: 
 \left( \frac{B}{10^9 {\rm G}} \right)^4
 \left( \frac{T}{10^7 {\rm K}} \right) \, .
\end{equation}           
The dependence $\varepsilon\propto B^4 T$ in the classical limit
could be expected from dimensional considerations.

The photon luminosity of the surface of a WD can be written as an 
effective emissivity $\eps_\gamma$, {\it i.e.\/} as an
energy-loss per volume and time,
\be
 \eps_\gamma = 3.3 \times 10^3 \: \frac{\rm erg}{\rm cm^{3}\: s} \: 
 \left( \frac{T}{10^7 {\rm K}} \right)^{7/2} \, .
\ee
If we require that a magnetized WD do not emit 
more energy in axions than in photons, we obtain
\begin{equation}  \label{wd,limit}
 g_{ae}\lsim 9 \cdot 10^{-13} \left( \frac{T}{10^7 {\rm K}} \right)^{5/4} 
                          \left( \frac{B}{10^{10} {\rm G}} \right)^{-2} \,.
\end{equation}
Since this bound is quite sensitive to $B$ and the knowledge of the
internal magnetic field strengths of WDs is poor, it is hard
to derive a precise bound for $g_{ae}$.

\section{Cooling times of magnetized white dwarfs}

The thermal history of a WD can be viewed essentially as a
cooling process \cite{sh83},
\be  \label{DGL}
 \frac{dt}{dT}=-\frac{c_V(T)}{\eps(T)} ,
\ee
where $T$ is the interior temperature which is assumed to be uniform
in the Mestel theory. Three different sources of emissivity $\eps$ become
important for a WD cooling down, $\eps = \eps_\gamma +\eps_\nu +\eps_a$. 
We neglect one of them, neutrino emissivity $\eps_\nu$, 
which is only for the hottest WDs with $T\gsim 10^{7.8}$K important
and decreases additionally the cooling times. Since
$\varepsilon_a \propto T$, compared to $\eps_\gamma \propto T^{7/2}$, 
axion emission becomes more and more
important during the cooling history of magnetized WDs. 
Therefore the fraction of magnetized WDs among all
WDs should diminish drastically for low enough luminosities.

In Table I, we compare the fraction of strongly magnetized WDs
with the fraction of all WDs in three different
temperature bins \cite{ra96,sc95}.
The distribution of hot strongly magnetized WDs adapted from
Ref. \cite{sc95} spans approximately three order of magnitudes in
surface dipole field strength, 
$B_{\rm dipole} = 3\cdot 10^6 - 10^9$ G, and  has a maximum at 
$B_{\rm dipole}\ap 3\cdot 10^7$ G. It consists of only 18 stars, so some
caution in interpreting the data is appropriate. Nevertheless, the
fraction of magnetized WDs in the last temperature bin is
considerably diminished compared to the total WD population.
One {\em possible\/} explanation for this could be the additional
energy loss of magnetized WDs due to axion cyclotron emission.

\begin{table}
\begin{center}
\footnotesize\rm
\caption{\label{tab}
Fraction of all white dwarfs (WDs) and of magnetized WDs in three
temperature bins.}
\begin{tabular}{ccc}               
\topline
 $T_{\rm eff} [10^3 $K] &  
 Fraction of all WDs & Fraction of magnetized WDs \\
\midline
 40-80 & 1\% & 0\% \\
 20-40 & 23\% & 50\% \\
 12-20 & 76\% & 50\% \\
\bottomline
\end{tabular}
\end{center}  
\end{table}

An obvious test for the hypothesis that magnetized WDs
 emit axions is the
comparison of the observed temperature distribution of disc magnetized
WDs  with the
predicted one. In Fig.~2 we show the calculated fraction of magnetized
WDs in
two different temperature bins as function of the magnetic field
strength. For comparison, the observed fractions of all disc WDs 
and of magnetized WDs  are shown  with
error bars on the left and on the right side of
the panel. For internal magnetic field strengths $B_{10}\lsim
10 (10 {\rm meV}/m_a)^2$, the observed temperature
distribution of all disc WDs is well reproduced, while for higher
magnetic field strengths the observed temperature distribution
of the magnetized WDs agrees also reasonably well with the predicted one.
($B_{10}$ denotes $B/10^{10}$~G.)

\begin{center}
\begin{figure}
\begin{minipage}[t]{3in}
\setlength{\unitlength}{1in}
\begin{picture}(6,0)
 \epsfig{file={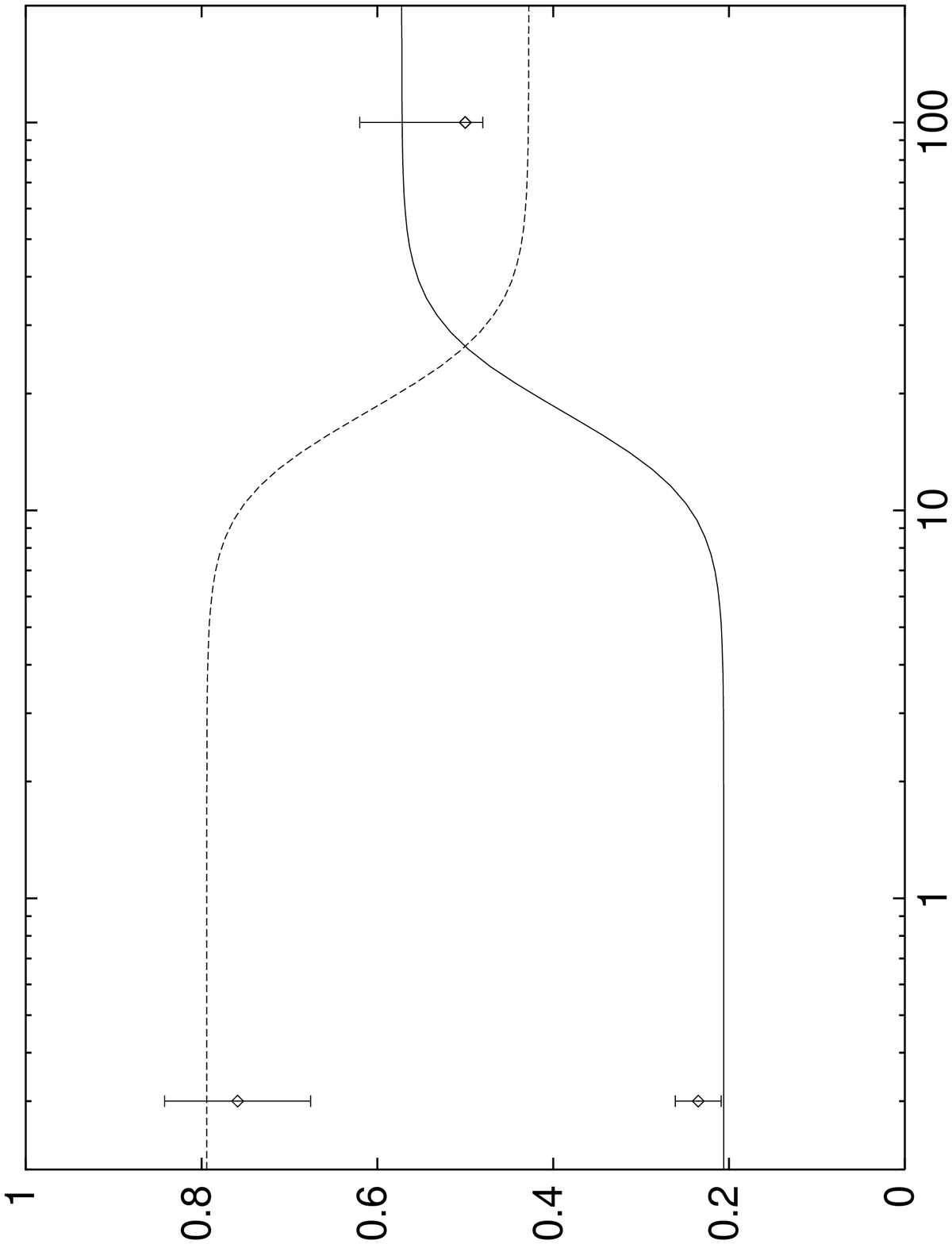}, height=3in, angle=270}
 \put(-3.05,-1.1){{\footnotesize $f_i$}}
 \put(-1.6,-2.35){{\footnotesize $B_{10}$}}
 \put(-2.5,-0.6){{\footnotesize $20000<T<40000$}}
 \put(-2.5,-1.5){{\footnotesize $12000<T<20000$}}
\end{picture}
\end{minipage}
 \vskip6.3cm
\caption{\label{frac}   
Fraction $f_i$ of magnetic WDs in two  temperature bins as function
of the magnetic field strength $B_{10}$ for $m_a=10$~meV and $n_{\rm
ion}=4.5\times 10^{28}$cm$^{-3}$; 
on the left side the observed values of all WD,
on the right side the observed values of magnetized WD.}
\end{figure}
\end{center}

The most stringent upper limit for the axion mass, $m_a\lsim 10$~meV,
was derived studying the evolution of red
giant stars. Using this value for the axion mass results in an average
internal field of $B_{10}\sim 30$ in order to reproduce the observed
temperature distribution of magnetized WDs. Although the internal field
strengths of magnetized white dwarfs 
are generally assumed to be in the range $B_{10}=0.01\ldots 1$,
they could approach even $B_{10}=100$ \cite{sh83}.
Nevertheless, if one concedes a factor 5 as error in the upper mass
limit, the necessary internal magnetic strengths of the
 magnetized white dwarfs  are in the general accepted range.

Finally, let us speculate about the nature of the galactic dark matter halo. 
A preferred interpretation of the microlensing events seen by MACHO and
EROS is that a substantial fraction of the galactic dark matter halo
consists of compact objects with the most probable mass around $0.5 M_\odot$.
For this mass range, WDs which are
known to exist in large numbers are the natural candidates,
although exotic objects like primordial black holes or gravitino stars
cannot be excluded.

Various optical searches \cite{suche}
did not detect a substantial halo population of WDs. Hence the
supposed halo population has to be dim enough to have eluded
detection. But then the estimated cooling times \cite{gr97}
of 11 to 15.5~Gyrs 
make them as old as globular clusters and raise the question whether
this allows reasonable life times for the progenitors of the WDs.
Therefore, such a halo seems to be only
possible if the halo WDs are distinguished by some physical
property which shortens their cooling times.
This different physical property could be a
strong internal magnetic field of the halo WDs. For magnetic
fields $B_{10}\gsim 5 \: ( 10 {\rm meV}/m_a)^{2}$, their cooling
times would drop below 1~Gyr and would make a halo 
consisting of magnetized WDs invisible.

\section{Summary}

We have derived the axion emissivity of a magnetized electron gas 
due to the process $e^- \to e^- +a$ for arbitrary magnetic 
field strengths $B$. 
Applying axion cyclotron emission as an additional cooling 
mechanism to neutron stars 
we could constrain the axion electron coupling constant to
$g_{ae} \lsim {\cal O} (10^{-10})$.
In the case of white dwarfs we could derive the more stringent limit
Eq. (\ref{wd,limit}). 
We have noted that the lack of low-temperature magnetized
white dwarfs could be interpreted as signature of an additional
energy loss due to axion cyclotron emission.

\section*{Acknowledgments}
I am grateful to C. Wilke and G. Wunner for 
collaboration on which this contribution is based on and to G. Raffelt
for discussions.
This work was supported by a Feodor-Lynen scholarship of the Alexander
von Humboldt-Stiftung.


\end{document}